# Superdirective Leaky Radiation from a PT-Synthetic Metachannel

Mehdi Hajizadegan and Pai-Yen Chen*

*Department of Electrical and Computer Engineering, University of Illinois at Chicago,*

*Chicago, IL 60661*

*Spectral singularities appearing in parity-time (PT)-symmetric non-Hermitian optical systems have aroused a growing interest due to their new, exhilarating applications, such as bifurcation effects at exceptional points [1]-[8] and the coexistence of coherent perfect absorber and laser (so-called CPAL point) [9]-[13]. We introduce here how the concept of CPAL action provoked in PT-symmetric metasurfaces can be translated into practical implementation of a low-loss zero- or low-index channel supporting a nearly undamped fast-wave propagation. Such a PT-synthetic metachannel shows the capability to produce a high-directivity leaky radiation, with a beam angle that can be altered by varying the gain-loss parameter. The proposed structure may enable new kinds of super-directivity antennas working in different regions of the electromagnetic spectrum, as well as various applications that demand extreme dielectric properties, such as epsilon-near-zero (ENZ).*

E-mail: pychen@uic.edu

Unusual points in the spectra of non-Hermitian physical systems, such as the exceptional point [1]-[8] and the merging point of laser and anti-laser (i.e., CPAL point) [9]-[13], have attracted substantial attention in the past few years. These isolated singular points are generally missed in a generic situation and may require special engineering of a quantum [14], optical [1]- [13], acoustic [15] or electronic [16],[17] systems. The existence of these points make the Taylor series expansion fail to describe the scattering function of waves. The CPAL systems with PT-symmetry is of particular interest because they allow simultaneous realization of laser and coherent perfect absorber (CPA) in a single component [9]-[13]. Traditionally, a laser oscillator emits coherent outgoing radiations, whereas a CPA is its time-reversed counterpart that represents a dark medium absorbing all incoming radiation. In general, both functions cannot coexist in a single device. Nonetheless, the continuum spectrum of non-Hermitian optical systems with PT-symmetry could host both spectral singularities at a given wavelength. At the CPAL point, the two eigenvalues of the scattering matrix tend to infinity and zero, corresponding to the laser and CPA modes, respectively. These two modes with dramatically different scattering responses can be switched by tuning the initial phase offset of two counter-propagating incoming beams. The ideal of CPAL action has been theoretically proposed [9]-[10] and later experimentally demonstrated [11]-[13] in waveguides and coupled resonators with spatially-distributed, balanced gain and loss. In this context, PT-symmetric metasurfaces have been recently proposed to make a simplified, low-profile alternative to CPAL devices [18], as well as negative-index media [19], optical sensing and imaging devices [20],[21], and the unidirectional reflectionless channel biased at the exceptional point [22].

In the Letter, we will propose new types of electromagnetic medium formed by PT-symmetric metasurfaces operating at the CPAL point. This system is composed of a pair of active and passive metasurfaces, with the spatial dependence of surface impedance given by $Z_s(y) = -R \cdot \delta(y - d/2) + R \cdot \delta(y + d/2)$ [see Fig. 1(a)], where $\pm R$ is the surface resistance, $d$ is the spacing between two metasurfaces, and $\delta$ is the Kronecker delta function. The spatially-distributed balanced gain $(-R)$ and loss $(+R)$ form the basis of a PT-symmetric system [1].

While scattering from *PT*-symmetric metasurfaces has been studied for different purposes, wave propagation and leakage in a *PT*-synthetic metachannel sketched in Fig. 1(a) (e.g., when excited by a waveguide port or point source) is yet to be explored. Understanding waveguiding characteristics and effective medium properties of a PT-synthetic metachannel may bring out new physical phenomena and applications underlying them. In the following, we will demonstrate that such a low-profile and relatively unsophisticated metachannel can exhibit extreme effective dielectric properties, such as epsilon-near-zero (ENZ), existing in a dispersive medium [6],[7] or waveguide operating at its cutoff frequency [23],[24]. More interestingly, by varying the dimensionless gain-loss parameter or non-Hermiticity, $\gamma = R/\eta$ ($\eta = \sqrt{\mu/\varepsilon}$ is the impedance of the background medium), the propagation constant of the guided transverse electric (TE) mode can be continuously tuned from zero to the background wavenumber ($k = \omega\sqrt{\mu\varepsilon}$). From the perspective of effective medium theory [23],[24], the effective permittivity of the PT-synthetic metachannel can vary from ENZ to that of the background medium (i.e., $0 \leq \text{Re}[\varepsilon_{eff}] \leq \varepsilon$). Unlike conventional ENZ and low-index media, the calculated $\text{Im}[\varepsilon_{eff}]$ related to the power attenuation rate or path loss could be vanishingly small. As a result, the almost undamped fast wave propagating in the PT-synthetic metachannel will produce coherent radiation leakage and form a superdirective radiation pattern, owning to an uniform and large radiating aperture. Additionally, by adjusting the gain-loss parameter ($\gamma$), the radiating angle can be reconfigured to any direction between broadside and end-fire.

In order to understand singularities in PT-symmetric metasurfaces, we first consider scattering of the TE-polarized plane wave from this composite structure [see Fig. 1(b)], which can be described by the two-port transmission-line network (TLN) in the inset of Fig. 1(b). In the TLN model, the background medium has a tangential wavenumber and a characteristic impedance given by $k_y = k\cos\alpha$ and $Z = \eta k / k_y$, respectively. The outgoing scattered waves and the incoming waves can be related by the scattering matrix, $\mathbf{S} = \begin{pmatrix} t & r^+ \\ r^- & t \end{pmatrix}$, where $t$ and $r$ are transmission and reflection coefficients for bottom ($-$) and top ($+$) incidences [25]. Figure 1(c) shows contours of two eigenvalues of $\mathbf{S}$ as a function of $\gamma$ and the angle of incidence $\alpha$, with the electrical length between the two metasurfaces $\Phi = k_y d = \pi/2$. We find that at the CPAL point, which exists when $\gamma = 1/\sqrt{2}\cos\alpha$ and $\Phi = \pi/2$, the two eigenvalues are zero and infinity. The exceptional point is also observed in Fig. 1(c). The two eigenvalues coalesce at this branch point singularity, dividing the system into the exact symmetry phase with unimodular eigenvalues and the broken symmetry phases with non-unimodular ones [10]. The CPAL point occurs in the broken symmetry phase, as is in most PT optical systems.

Next, we will discuss the use of PT-symmetric metasurfaces as a guided propagation channel and will show that the CPAL point found in scattering events can shed light on tailoring effective medium properties of a PT-synthetic metachannel. The eigenmodal solutions of a PT-synthetic metachannel can be derived from the transverse-resonance relation that considers an equivalent TLN model similar to that used for the scattering event [the inset of Fig. 1(b)] [26],[27]; here, the line has a transverse propagation constant $k_y = \sqrt{k^2 - \beta^2}$ and a characteristic impedance for the TE mode given by $Z = \eta k / k_y$, where $\beta$ is the longitudinal propagation constant. The transverse resonance condition means that at any point along the $y$-axis (transverse direction), the sum of the input impedance looking to the $+\hat{y}$ side, $Z_{in}^{(+)}$, and that looking into the $-\hat{y}$ side, $Z_{in}^{(-)}$, must be zero. This yields a dispersion equation given by:

$$\tan\left(\sqrt{k^2-\beta^2}\,d\right) = \frac{j(k^2-\beta^2)}{(k^2-\beta^2)-\mu^2\omega^2/2R^2}. \qquad (1)$$

When the CPAL condition is met (i.e., $\gamma = 1/\sqrt{2}\cos\alpha$ and $\Phi = \pi/2$), solving Eq. (1) leads to a purely real propagation constant given by $\beta = k\sin\alpha$. Interestingly, the seemingly unrelated scattering and guided propagation problems can be correlated at the singular point. In the scattering event [Fig. 1(b)], the laser mode of CPAL action is achieved when $Z_{in}^{(+)} + Z_{in}^{(-)} = 0$ at any point along the $y$-axis, such that the scattering coefficients in **S** become infinite. Since the scattering and guided propagation problems share a similar TLN model shown in Fig. 1(b) (although $Z$ and $k_y$ are defined differently), if the system is locked at the CPAL point, the longitudinal propagation constant, $\beta$, is identical to the tangential wavenumber in the scattering event, $k_x = k\sin\alpha$. This result is consistent with the eigenmodal solution obtained from Eq. (1).

We note further that a PT-synthetic metachannel locked at the CPAL point exhibits a fast-wave propagation behavior (i.e., $\beta < k$) and, thus, has a low effective permittivity given by $\varepsilon_{eff}/\varepsilon = \sin^2\alpha$. Fast waves propagating in the unbounded PT channel corresponds to the leaky-wave mode [27],[28], which will induce radiation leakage with a beam angle measured from broadside as $\alpha = \sin^{-1}(\beta/k)$. More interestingly, the gain-loss parameter governs the radiation direction, analogous to how it controls the CPAL action at a certain angle of incidence, $\alpha$, in the scattering event. We first consider a metachannel composed of PT-symmetric metasurfaces with $\gamma = 1/\sqrt{2}$ and height of one-quarter wavelength, which makes a CPA-laser for normally-incident waves at frequency $f_0$. Based on the above discussions, when a PT-synthetic metachannel is excited by a waveguide port at $f_0$, one can expect that $\beta = 0$ and, thus, an ENZ medium with infinite phase velocity is achieved. Figure 2(a) and 2(b) show the calculated radiation pattern [25] and electrical field distributions [29] for this unbounded PT channel at $f = f_0 - \delta f$; here, $\delta f = 10^{-4} f_0$ that leads to $\beta / k = 0.005 - j0.010$. It is seen from Fig. 2(b) that inside the PT channel, a nearly constant phase distribution can be obtained due to the peculiar ENZ characteristics. Moreover, the nearly undamped fast-wave property with $\beta \sim 0$ results in a highly directive broadside radiation, as can be seen in Fig. 2(a). In the far (Fraunhofer) zone, the directivity of 2-D radiative apertures can be defined quantitively as the ratio of the maximum radiation intensity of the main lobe ($U_{max}$) to the average radiation intensity over all space [28]:

$$D_{max} = \frac{U_{max}}{P_{rad}/2\pi} = \frac{2\pi |E_z(\alpha)|^2}{\int_{-\pi}^{\pi} |E_z(\theta)|^2 d\theta}, \qquad (2)$$

where $P_{rad}$ is the total radiated power. Our calculations show that the directivity of beam increases with increasing the channel length $L$. For example, $D_{max}$ is 10.89 (10.37 dB) for $L = 2\lambda$, and is increased to 51.86 (17.15 dB) for $L = 10\lambda$, and 138.84 (21.43 dB) for an infinitely long structure. Given that $\text{Im}[\beta] \sim 0$, in light of the contactless gain-loss interaction, the PT leaky-wave structure can have a very large effective aperture [25] and superdirectivity. Moreover, changing the gain-loss parameter will alter the beam angle, as can be seen in far-field radiation patterns in Fig. 2(c) and contour plots of electric field distributions in Fig. 2(b). For different targeted beam angles $\alpha = 0°$, 30°, 45°, and 60°, surface resistances of two metasurfaces and the spacing between them must be changed accordingly ($\gamma = 1/\sqrt{2} \cos\alpha$), in order to lock the system at the CPAL point. The radiation pattern is somehow bidirectional, as a result of unidirectional scattering responses of PT systems [22]. Compared with other ENZ medium made of metamaterials or Drude-dispersion materials, the proposed low-index metachannel may not only ease manufacturing complexity, but also considerably reduce the attenuation rate, thus facilitating the practice of ENZ-allowed applications (e.g., supercoupling and superluminal effect, energy squeezing, and enhanced nonlinear wave mixing [6],[7],[23],[24], as well as leaky-wave radiations [7],[27]). Leaky-wave antennas based on guided-wave channels with periodic grids or slots have been enormously studied in different spectral ranges. However, the non-negligible attenuation rate basically limits the effective aperture of these leaky-wave structures, particularly for optical applications. Besides, the occurrence of higher-order (Floquet) spatial harmonics could produce unintended grating lobes. These long-standing challenges may be addressed by the PT-synthetic leaky-wave structures with homogeneous surface impedances and contactless gain-loss interactions.

We also analyze radiation from an electric line source ($\bar{J} = \hat{z} I_0 \delta(x)\delta(y)$ [A/m$^2$]) located at the center of the PT-synthetic metachannel, as schematically shown in Fig. 3(a). Figures 3(b) and 3(c) show the far-field radiation pattern and contours of electric field distributions for the PT-synthetic metachannel in Fig. 2(c), under excitation of a line source; here, the operating frequency has an $-1\%$ offset from the CPAL point. The electric field in the far zone can be obtained as an inverse Fourier transform [28]:

$$E_z(x,y) = \frac{1}{2\pi} \int_{-\infty}^{+\infty} \tilde{E}_z(k_x) e^{-j(k_x x + k_y y)} dk_x \qquad (3)$$

where $\tilde{E}_z$ is the spectral electric field on the metasurface. From Fig. 3(b), we find that the agreement between analytical (lines) and numerical (dots) results is excellent, and that radiation from the line source can be reshaped into a directive beam and can be steered towards a specific direction in the far field. The beam angle that depends on the gain-loss parameter can be continuously tuned from broadside towards end-fire direction, as can be seen in Figs. 3(b) and 3(c). Our results demonstrate that a highly directive and reconfigurable antenna or emitter can be realized by exploiting the CPAL singularity, at which the transverse resonance relation is satisfied at any point of arbitrary cross sections of a metachannel [18]. Finally, we also briefly discuss the practical implementation of PT-symmetric metasurfaces. The positive surface resistance $R$ can be readily achieved by a resistive sheet or passive metasurface made of lossy materials. In the optical region, an active metasurface could be a (patterned) thin layer of material with negative conductivity (e.g., optically-pumped 2D materials [30], organic dyes, or semiconductors). The active metasurface working at microwave frequencies could be a metasurface loaded with negative-resistance elements [17],[31].

In conclusion, we have proposed the concept of a PT-synthetic metachannel exhibiting zero or low effective permittivity, for which the CPAL point offers a comprehensive guidance on tailoring the extreme effective permittivity. When this metachannel locked at the CPAL point is fed by a waveguide port or line source, the leaky-wave mode can couple the guided fast wave into the background medium, resulting in a highly directive radiation leakage. Additionally, the beam can be steered from broadside towards end-fire direction by controlling the gain-loss parameter. We envision that the proposed active component may be applied to many applications of interest in different electromagnetic spectra, including the high-directivity antenna or emitter with tunable radiating angles, as well as low-attenuation ENZ or low-index media.

This work was partially supported by NSF ECCS-CCSS CAREER No.1914420.

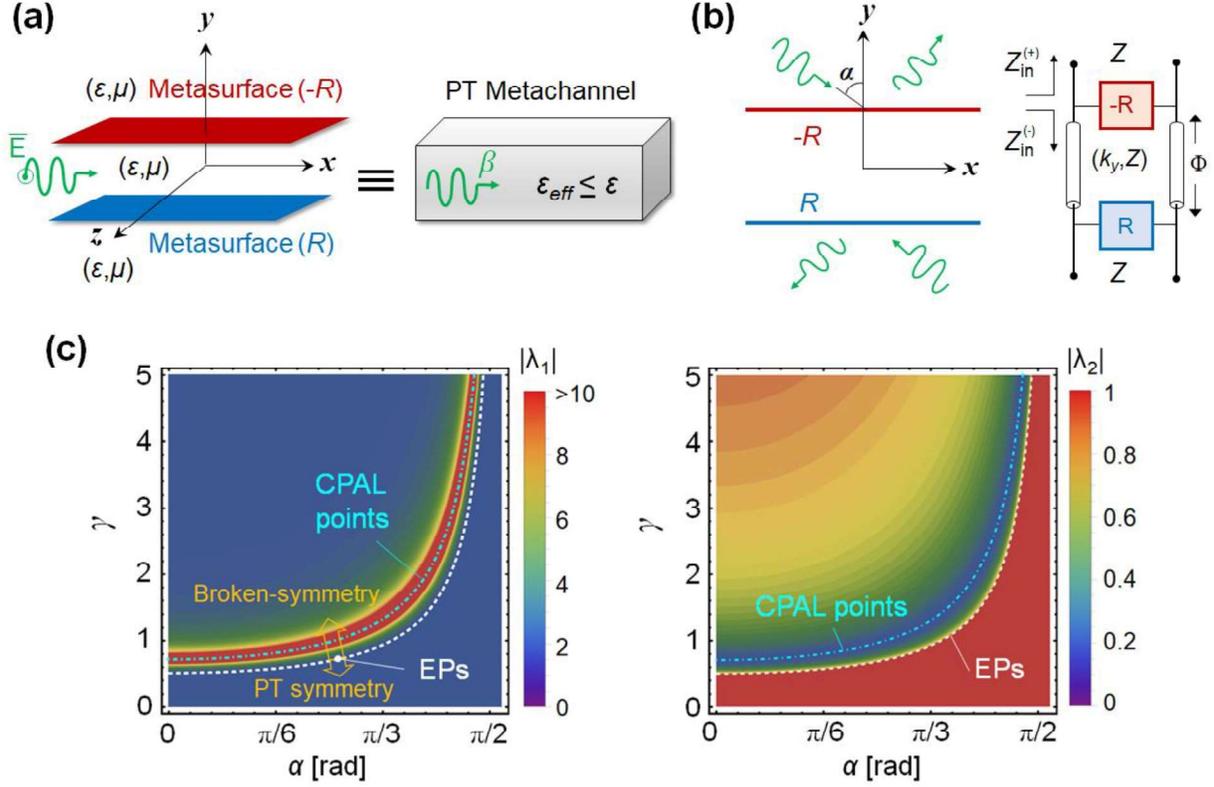

**Figure 1.** (a) Schematics of PT-synthetic metachannel composed of an active metasurface (-*R*) and a passive metasurface (*R*). The metachannel has a longitudinal propagation constant *β* that can be varied between zero and the wavenumber of the background medium, corresponding to an effective permittivity, $0 \leq \varepsilon_{eff} \leq \varepsilon$. (b) Scattering of plane waves by PT-symmetric metasurfaces and its corresponding transmission-line network (TLN) model. (c) Contours of two eigenvalues of the scattering matrix for the PT scattering system in (b), as a function of the gain-loss parameter *γ* and the angle of incidence *α*.

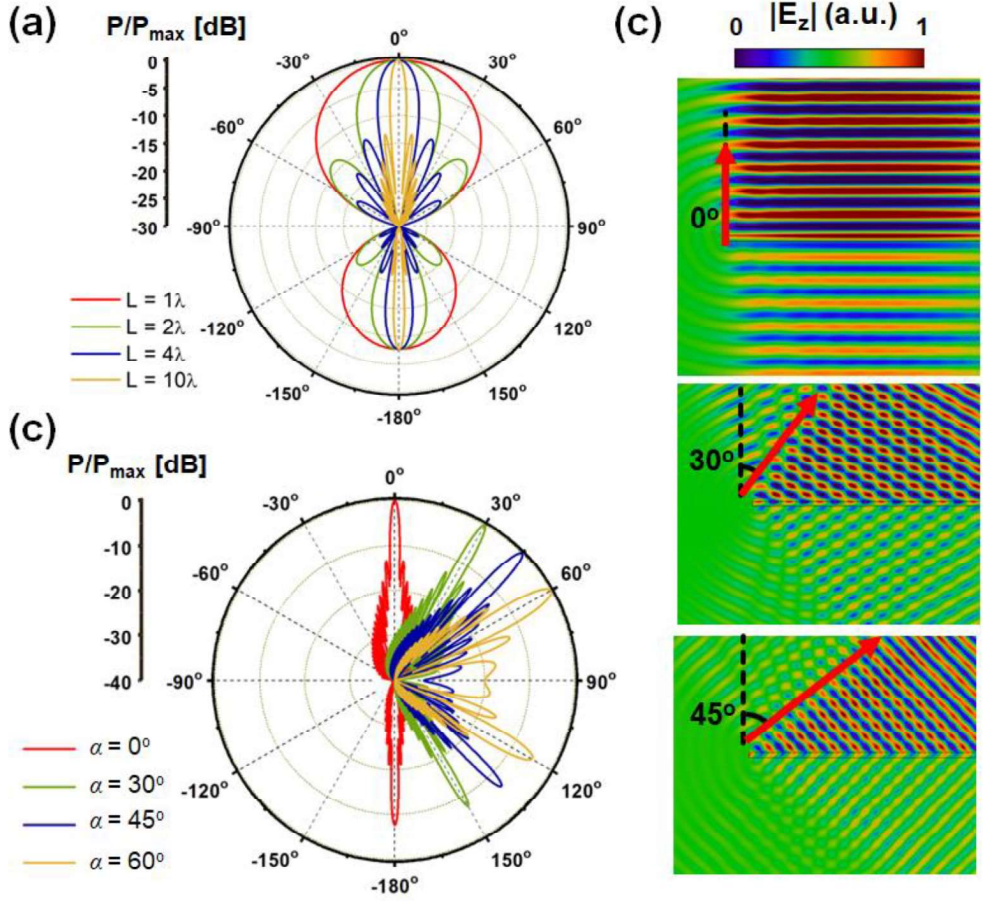

**Figure 2.** (a) Far-field (Fraunhofer) radiation patterns for PT metachannels with $\gamma = 1/\sqrt{2}$, $\Phi = \pi/2$, and different lengths $L$. The structure is excited by a waveguide port on the left. Due to the channel's ENZ characteristics, broadside radiations are observed. (b) Snapshots of electric field distributions for the PT metachannels, which can steer a beam (leaky wave) to different angles by changing $\gamma$. (c) Radiation patterns for CPAL-locked PT metachannels with $\gamma = 1/\sqrt{2} \cos\alpha$ and $\Phi = \pi/2$, where $\alpha$ is the beam angle; in all cases, the channel length is fixed to 20 $\lambda$. Their corresponding snapshots of electric field distributions can be seen in (b).

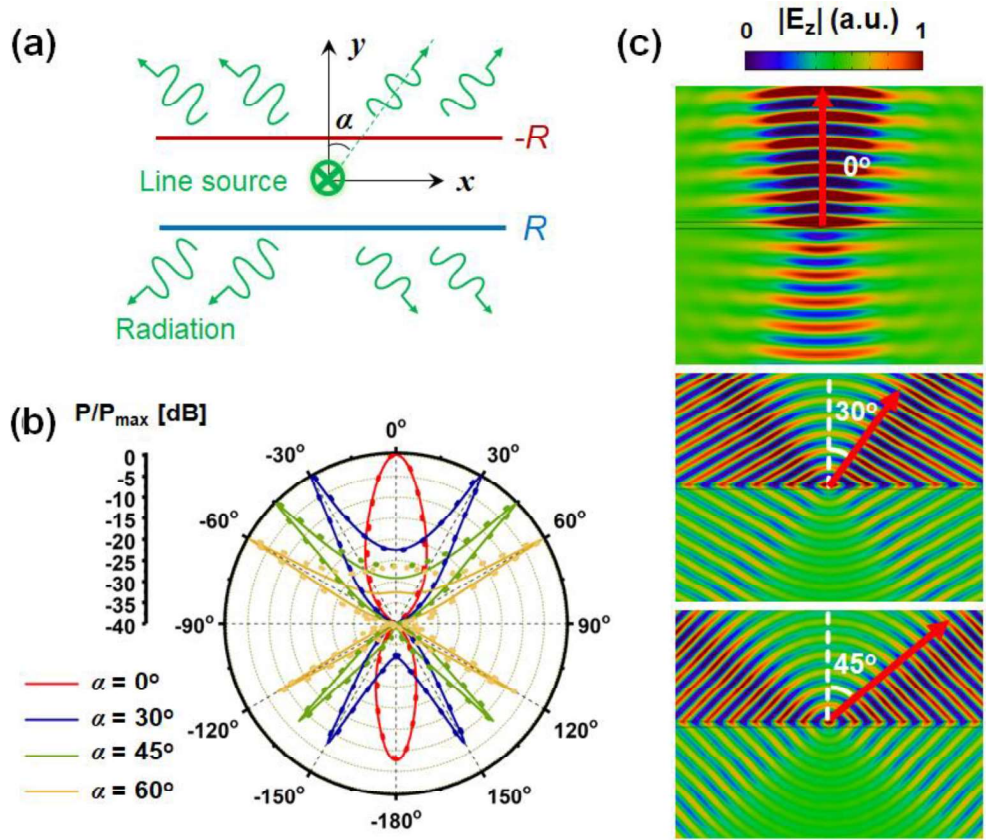

**Figure 3.** (a) Schematics, (b) far-field radiation patterns, and (c) snapshots of electric field distributions for CPAL-locked PT metachannels with $\Phi = \pi/2$, excited by a line source; for a certain beam angle $\alpha$, the corresponding gain-loss parameter $\gamma = 1/\sqrt{2} \cos\alpha$.

# Superdirective Leaky Radiation from a PT-Synthetic Metachannel

## S1. Coherent perfect absorber-laser using PT-symmetric metasurface

Consider scattering of a plane wave (wave vector $\mathbf{k}=\hat{x}k_x+\hat{y}k_y$) from PT-symmetric metasurfaces, the elements of the scattering matrix can be calculated using the two-port transmission-line network (TLN) model in Fig. 1(b). The background medium has a characteristic impedance $Z$, and the two shunt surface resistances are separated by a portion of transmission line with a characteristic impedance $Z$ and an electrical length $\Phi=k_y d$. The surface resistances have opposite values $\pm R$. In this system, the dimensionless gain-loss parameter (or non-Hermiticity parameter) can be defined as $\gamma = R/Z$. For the transverse electric (TE) plane wave incident at an arbitrary angle $\alpha$, the wave impedance is given by $Z=\eta/\cos\alpha$ and the propagation constant is $k_y = k\cos\alpha$, where $\eta$ is the characteristic impedance of background medium. Similar considerations apply to the transverse magnetic (TM) incidence, but with $Z=\eta\cos\alpha$. Using the transfer matrix formalism, and assuming time-harmonic fields $e^{j\omega t}$, the scattering parameters, involving transmission ($t$) and reflection ($r$) coefficients for bottom ($-$) and top ($+$) incidences are obtained as:

$$\mathbf{S} = \begin{pmatrix} t & r^+ \\ r^- & t \end{pmatrix}$$

$$= \begin{pmatrix} \dfrac{1}{e^{jx} - j\sin(x)/(2\gamma^2)} & \dfrac{(1+2\gamma)\sin(x)}{-\sin(x)-2j\gamma^2 e^{jx}} \\ \dfrac{(1-2\gamma)\sin(x)}{-\sin(x)-2j\gamma^2 e^{jx}} & \dfrac{1}{e^{jx} - j\sin(x)/(2\gamma^2)} \end{pmatrix}. \qquad (S1)$$

The validity of PT-symmetry imposes a generalized conservation relation on the scattering matrix: $\mathbf{S}^*(\omega) = \mathcal{PT}\mathbf{S}(\omega)\mathcal{PT} = \mathbf{S}^{-1}(\omega)$ [1],[2], where the parity operator $\mathcal{P} = \begin{pmatrix} 0 & 1 \\ 1 & 0 \end{pmatrix}$, the time-reversal operator $\mathcal{T} = \begin{pmatrix} 0 & 1 \\ 1 & 0 \end{pmatrix}\mathcal{K}$, and $\mathcal{K}$ is the complex conjugation operator.

A condition of special interest resides in the exceptional point, when $\gamma = 1/(2\cos\alpha)$, the unidirectional reflectionless propagation can be achieved. In addition to this branch point singularity, the CPAL action is achieved when $\gamma = 1/(\sqrt{2}\cos\alpha)$ and $\Phi = \pi/2$. Electric fields on bottom ($-$) and top ($+$) sides can be decomposed into forward ($f$)- and backward ($b$)-propagating waves, whose relations can be described by the transfer matrix $\mathbf{M}$ as: $\begin{pmatrix} E_f^+ \\ E_b^+ \end{pmatrix} = \mathbf{M} \begin{pmatrix} E_f^- \\ E_b^- \end{pmatrix}$. The CPAL system based on PT-symmetric metasurfaces can operate in the laser mode when $E_b^+/E_f^- \neq M_{21}$, or in the CPA mode when $E_b^+/E_f^- = M_{21}$. The lasing oscillator mode provides output fields $E_b^-, E_f^+ \neq 0$ even for zero input fields ($E_f^-, E_b^+ \approx 0$), while the CPA mode makes $E_b^- = E_f^+ = 0$ even for non-zero input fields ($E_f^-, E_b^+ \neq 0$) [3][4].

**S2. Eigenmodes in a PT-synthetic channel**

Consider first the eigenmodes of the PT-synthetic channel in Fig. 1(a), a guided wave propagates along the $x$-axis with a factor $e^{-j\beta x}$. Electromagnetic fields can be separated into transverse electric (TE) and transverse magnetic (TM) fields with respect to a lateral coordinate. The TE mode has the following electric field distributions:

$$\mathbf{E} = \hat{\mathbf{z}} \, E_z(y) e^{j(\omega t - \beta x)}$$

$$E_z(y) = \begin{cases} c_1^{TE} e^{-j\sqrt{k^2-\beta^2}(y-d/2)} & \text{if } y \geq d/2 \\ c_2^{TE} e^{j\sqrt{k^2-\beta^2}y} + c_3^{TE} e^{-j\sqrt{k^2-\beta^2}y} & \text{if } -d/2 \leq y \leq d/2 \\ c_4^{TE} e^{j\sqrt{k^2-\beta^2}(y+d/2)} & \text{if } y \leq -d/2 \end{cases} \quad (S2)$$

where $\beta$ is the (longitudinal) propagation constant, the transverse propagation constant $k_y = \sqrt{k^2 - \beta^2}$, $k = \omega\sqrt{\mu\varepsilon}$, $\omega$ is the angular frequency, $\mu$ and $\varepsilon$ are the wavenumber, permeability and permittivity of the background medium, respectively. Electric and magnetic fields (**E**, **H**) for the TE mode in each region can be obtained from source-free Maxwell's equations. The complex coefficients $c_i^{TE}$ may be determined by matching the boundary conditions enforced on the metasurface: $\mathbf{J}_s = \hat{\mathbf{n}} \times (\mathbf{H}^+ - \mathbf{H}^-) = \mathbf{E}_{tan}/Z_s$ and $\hat{\mathbf{n}} \times (\mathbf{E}^+ - \mathbf{E}^-) = 0$, where $\hat{\mathbf{n}}$ is the surface normal vector, and the surface impedance has a PT-symmetric profile: $Z_s(y) = -R \cdot \delta(y - d/2) + R \cdot \delta(y + d/2)$, $R$ and $-R$ are the surface resistances for the passive and active metasurfaces, respectively. The resulting dispersion equation for the complex eigenmodal solution $\beta$ is given by:

$$\tan\left(\sqrt{k^2 - \beta^2}\, d\right) = \frac{j(k^2 - \beta^2)}{(k^2 - \beta^2) - \omega^2 \mu^2 / 2R^2}. \quad (S3)$$

The dispersion equation can also be solved by using the transverse resonance technique [4], which employs a transmission-line model of the transverse cross section of a waveguiding structure. Eigenmodes are obtained in a resonant line, if the sum of the input impedances seen looking to either size of an arbitrary point $y'$ is zero, namely:

$$Z_{in}^{(+)}(y') + Z_{in}^{(-)}(y') = 0. \quad (S4)$$

To find the eigenmodes for the TE mode, the equivalent transverse resonance circuit shown in the inset of Fig. 1(b) can be used. The line for $-d/2 \leq y \leq d/2$ represents the PT-synthetic channel and has a transverse propagation constant $k_t$ and a characteristic impedance for TE modes given by $Z = k\eta/k_y$ and $\eta = \sqrt{\mu/\varepsilon}$. Due to the fact that the longitudinal propagation constant, $\beta$, must be the same in both regions for phase matching of the tangential fields at the interface. For $y < -d/2$ and $y > d/2$, the transverse line is terminated with an impedance given by $Z = k\eta/k_y$. Applying the transverse resonance condition (Eq. S4) will lead to the dispersion equation in Eq. S3.

### S3. Radiation from a PT-synthetic channel under excitation of a waveguide port

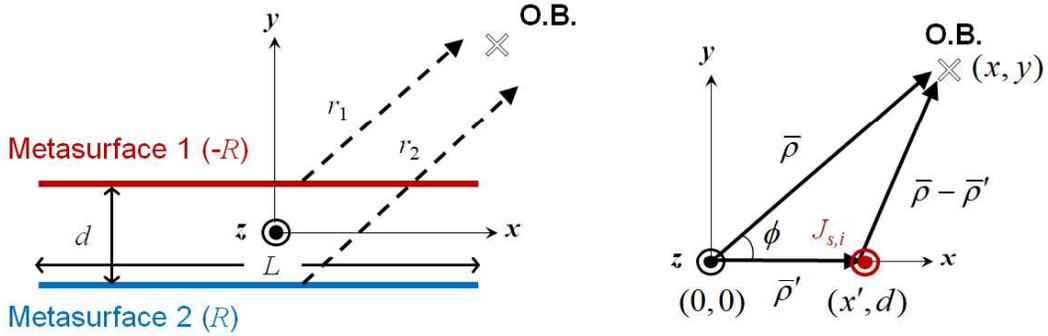

**Figure S1.** PT-synthetic metachannel geometry and far-field approximations.

The electric surface current density on a metasurface is induced by discontinuity of magnetic fields. For the PT-symmetric metasurface channel sketched in Fig. S1, surface current densities are given by:

$$\mathbf{J}_{s,1} = \hat{z} J_{z,1} = \hat{y} \times [\mathbf{H}_+ - \mathbf{H}_-]\big|_{y=d/2} = \frac{\hat{z}\left(E_z\big|_{y=d/2}\right)}{-R};$$

$$\mathbf{J}_{s,2} = \hat{z} J_{z,2} = -\hat{y} \times [\mathbf{H}_+ - \mathbf{H}_-]\big|_{y=-d/2} = \frac{\hat{z}\left(E_z\big|_{y=-d/2}\right)}{R}. \tag{S5}$$

In the far-field (Fraunhofer) region, the electric and magnetic fields due to $\mathbf{J}_s$ is given by

$$\mathbf{E} = -j\omega\mathbf{A} + \frac{1}{j\omega\varepsilon\mu}\nabla(\nabla\cdot\mathbf{A}) \approx -j\omega\mathbf{A},$$
$$\mathbf{H} = \frac{1}{\mu}\nabla\times\mathbf{A},$$
(S6)

where the magnetic vector potential $\mathbf{A}$ due to $\mathbf{J}_s$ is given in terms of the Green's function:

$\mathbf{A}(\bar{\rho}) = \iint_{s'} \mathbf{J}_s \frac{\mu}{4j} H_0^{(2)}(k|\bar{\rho}-\bar{\rho}'|) ds'$, where $g(\bar{\rho},\bar{\rho}') = \frac{1}{4j} H_0^{(2)}(k|\bar{\rho}-\bar{\rho}'|)$ is the two-dimensional Green's function, $\bar{\rho}' = x'\hat{x} + y'\hat{y}$ and $\bar{\rho} = \rho\hat{\rho} = x\hat{x} + y\hat{y}$ ($\rho$ is the radial distance and $\hat{\rho} = \cos\phi\hat{x} + \sin\phi\hat{y}$) are the position vectors of the source and the observer, respectively, and $H_0^{(2)}(\cdot)$ is the Hankel function of the second kind. In the far zone, the electric and magnetic fields produced by sheet currents induced on the metasurfaces only have $\hat{z}$ and $\hat{\phi}$ components in the cylindrical coordinates. Those constitute a transverse electromagnetic (TEM) wave propagating in the $\hat{\rho}$-direction, given by:

$$\mathbf{E} = \hat{z}E_z \simeq -\hat{z}j\omega A_z;$$
$$\mathbf{H} = \hat{\phi}H_\phi \simeq \hat{\phi} E_z/\eta.$$
(S7)

The time-averaged Poynting vector is therefore written as:

$$\mathbf{W} = \frac{1}{2}\text{Re}\left[\hat{z}E_z \times \hat{\phi}H_\phi^*\right] = \hat{\rho}\frac{1}{2\eta}|E_z|^2.$$
(S8)

Approximations can be made, especially for the far-field region that is usually the one of most practical interest, to simplify the formulation of fields radiated by a PT metachannel with length $L$ and infinite width, as sketched in Fig. S1(a). In the far zone, the distance from any point on the active metasurface ($(x',d/2)$ for $-L/2 \leq x' \leq L/2$) to the observation point can be approximately expressed as:

$$r_1 = |\bar{\rho} - \bar{\rho}'| = \sqrt{(x-x')^2 + (y-d/2)^2}$$
$$= \sqrt{\rho^2 - 2x'\rho\cos\phi - \rho d\sin\phi + (x')^2 + (d/2)^2} \qquad (S9)$$
$$\approx \rho - x'\cos\phi - d\sin\phi/2 \quad \text{for the phase term in the far zone}$$
$$\approx \rho \quad \text{for the amplitude term in the far zone,}$$

where $\phi$ is the angle in cylindrical coordinates measured from the x-axis [Fig. S1(b)]. Similarly, the distance from any point on the passive metasurface ($(x', -d/2)$ for $-L/2 \leq x' \leq L/2$) to the observation point can be written as:

$$r_2 = |\bar{\rho} - \bar{\rho}''| = \sqrt{(x-x')^2 + (y+d/2)^2}$$
$$\approx \rho - x'\cos\phi + d\sin\phi/2 \qquad (S10)$$
$$\approx \rho.$$

In the far zone where $\rho \gg \rho'$ and $k\rho \gg 1$, the vector potentials for the two metasurfaces are approximately given by:

$$\mathbf{A}_1 = \hat{\mathbf{z}} A_{z,1} \approx \hat{\mathbf{z}} \frac{\mu}{4j} \int_{-L/2}^{L/2} J_{z,1}(x') \frac{e^{-jk\rho}}{\sqrt{8j\pi k\rho}} e^{j(k\cos\phi \cdot x' + k\sin\phi \cdot d/2)} dx';$$
$$\mathbf{A}_2 = \hat{\mathbf{z}} A_{z,2} \approx \hat{\mathbf{z}} \frac{\mu}{4j} \int_{-L/2}^{L/2} J_{z,2}(x') \frac{e^{-jk\rho}}{\sqrt{8j\pi k\rho}} e^{j(k\cos\phi \cdot x' - k\sin\phi \cdot d/2)} dx'. \qquad (S11)$$

Since Maxwell's equations are linear, superposition applies and, therefore, the electromagnetic fields produced by the two currents sheets induced on the active and passive metasurfaces can be expressed as $\mathbf{E} = \mathbf{E}_1 + \mathbf{E}_2$, where $\mathbf{E}_1 \approx -\hat{\mathbf{z}} j\omega A_{z,1}$ and $\mathbf{E}_2 \approx -\hat{\mathbf{z}} j\omega A_{z,2}$ are electric fields produced by $\mathbf{J}_{s,1}$ and $\mathbf{J}_{s,2}$, respectively.

## S4. Radiation from a PT-synthetic channels under excitation of an electric line source

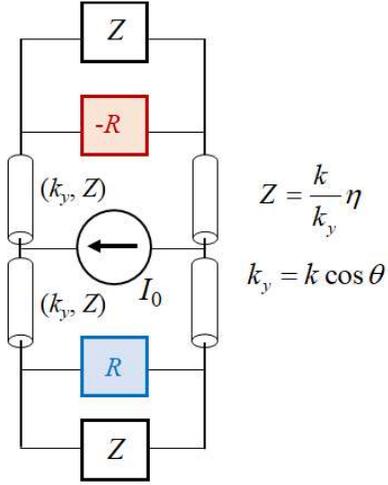

**Figure S2.** Transmission-line network model for the PT metachannel excited by a line source.

The structure considered here is PT-symmetric metasurfaces excited by an electric line source along $z$ with a time-harmonic dependence, embedded in the middle of the *PT*-symmetric metasurfaces, as sketched in Fig. 3(a). The transverse-equivalent network [6]-[9] as in Fig. S2 can be used to model such an antenna. The electric field in the background produced by a unit amplitude electric line source can be represented as an inverse Fourier transform,

$$E_z(x,y) = \frac{1}{2\pi} \int_{-\infty}^{+\infty} \tilde{E}_z(k_x) e^{-j(k_x x + k_y y)} dk_x, \qquad (S12)$$

where $\text{Im}[k_y] \leq 0$ in order to satisfy the radiation condition at infinity. The spectral electric field at the background-active metasurface interface is given by:

$$\tilde{E}_z^{(+)}(k_x) = \frac{e^{j\frac{\beta d}{2}} RZ \left[ Z - e^{j\beta d}(Z+2R) \right]}{e^{j2\beta d}(Z^2 - 4R^2) - Z^2}. \qquad (S13)$$

Similarly, the spectral electric field at the background-passive metasurface interface is:

$$\tilde{E}_z^{(-)}(k_x) = \frac{e^{j\frac{\beta d}{2}} ZR\left[e^{j\beta d}(Z-2R)-Z\right]}{e^{j2\beta d}(Z^2-4R^2)-Z^2}. \tag{S14}$$

The characteristic impedance, Z, for the TE and TM polarizations have the following expressions:

$$\begin{aligned} Z^{TE} &= (k/k_y)\eta; \\ Z^{TM} &= (k_y/k)\eta, \end{aligned} \tag{S15}$$

where the vertical wavenumber $k_y$ depends on the spherical angle $\theta$ as $k_y = \sqrt{k^2 - k_x^2} = k\cos\theta$. The TLN model can be used for determination of fields radiated by a source through an application of the reciprocity theorem. In this case, The far-zone electric field can readily be obtained through an asymptotic evaluation of Eq. S12 for large distances from the origin (i.e., $\rho \gg d$) [8]-[9]. The result for the upper half-plane is given by:

$$E_z^{(+)}(\rho,\theta) = E_z^{ff,(+)}(\theta) e^{-jk\rho}/\sqrt{\rho}, \tag{S16}$$

where the normalized far-field pattern is

$$E_z^{ff,(+)}(\theta) = \cos\theta \sqrt{\frac{jk}{2\pi}} \tilde{E}_z^{(+)}(k\sin\theta), \tag{S17}$$

and $\theta$ is the angle measured from broadside, and $\pm$ represents the upper and lower half-planes. Similarly, the far-zone electric field in the lower half-plane is given by:

$$E_z^{(-)}(\rho,\theta) = E_z^{ff,(-)}(\theta) e^{+jk\rho}/\sqrt{\rho}, \tag{S18}$$

where

$$E_z^{ff,(-)}(\theta) = \cos\theta \sqrt{\frac{jk}{2\pi}} \tilde{E}_z^{(-)}(k\sin\theta). \tag{S19}$$

The radiated power density in the upper and lower half-planes are given by:

$$P^{(\pm)}(\theta) = \frac{|E_z^{ff,(\pm)}(\theta)|^2}{2\eta}. \tag{S20}$$

The maximum radiated power is obtained if the slab thickness is equal to one quarter-wavelength.